\documentstyle[11pt]{article}
\def\subsubsection{\@startsection{subsubsection}{3}{\z@}{-3.25ex plus
 -1ex minus -.2ex}{1.5ex plus .2ex}{\large\sc}}

\newcommand{\be}{\begin{equation}}
\newcommand{\bel}[1]{\begin{equation}\label{#1}}
\newcommand{\ee}{\end{equation}}
\newcommand{\bea}{\begin{eqnarray}}
\newcommand{\ba}{\begin{array}}
\newcommand{\eea}{\end{eqnarray}}
\newcommand{\ea}{\end{array}}

\newcommand{\ra}{\rightarrow}

\newcommand{\disp}{\displaystyle}

\newcommand{\hsix}{\hspace*{6mm}}
\newcommand{\hfour}{\hspace*{4mm}}

\newcommand{\vfour}{\vspace*{4mm}}
\newcommand{\vtwo}{\vspace*{2mm}}
\newcommand{\htwo}{\hspace*{2mm}}
\newcommand{\honecm}{\hspace*{1cm}}

\newcommand{\udl}{\underline}

\newcommand{\ket}[1]{\mbox{$| \, {#1}\, \rangle$}}
\newcommand{\exval}[1]{\mbox{$\langle \, {#1}\, \rangle$}}

\begin{document}

\begin{titlepage}
\thispagestyle{empty}
\begin{center}
\vspace*{1cm}
{\large \bf
Reaction-Diffusion Processes of Hard-Core Particles
}\\[25mm]

{\large {\sc
Gunter M. Sch\"utz}
}\\[8mm]

\begin{minipage}[t]{13cm}
\begin{center}
{\small\sl
Department of Physics\\
University of Oxford\\
Theoretical Physics, 1 Keble Road, Oxford OX1 3NP, UK
}
\end{center}
\end{minipage}
\vspace{30mm}
\end{center}
{\small
We study a 12-parameter stochastic process involving particles with two-site
interaction and hard-core repulsion on a $d$-dimensional lattice.
In this model, which includes the asymmetric exclusion process, contact
processes and other processes, the stochastic variables are
particle occupation numbers taking values $n_{\vec{x}}=0,1$.
We show that on a 10-parameter submanifold the $k$-point equal-time
correlation functions $\exval{n_{\vec{x}_1} \cdots n_{\vec{x}_k}}$
satisfy linear differential-difference equations involving no
higher correlators. In particular, the average density $\exval{n_{\vec{x}}} $
satisfies an integrable diffusion-type equation. These properties are
explained in terms of dual processes and various duality relations are derived.
By defining the time evolution
of the stochastic process in terms of a quantum Hamiltonian $H$,
the model becomes equivalent to a lattice model in thermal
equilibrium in $d+1$ dimensions. We show that the spectrum of
$H$ is identical to the spectrum of the quantum Hamiltonian of a
$d$-dimensional, anisotropic spin-1/2 Heisenberg model. In one
dimension our results hint at some new algebraic structure behind the
integrability of the system.
}
\\
\vspace{5mm}\\
\udl{Keywords:} Reaction-diffusion processes, correlation functions, integrable
models\\
\udl{PACS numbers:} 05.40+j, 05.70.Ln, 02.50.Ey, 75.10.Jm
\end{titlepage}
\newpage

\section{Introduction}
\setcounter{equation}{0}

An interesting class of non-equilibrium problems with a rich dynamical
behaviour are stochastic reaction-diffusion systems \cite{1,2}.
These processes may involve one or several species of particles
$A,B,C, \dots $ and an inert state $\emptyset$ equivalent to
the absence of any of the interacting particles. Examples of such processes
are, to name but a few, coagulation $A+A \ra A$ \cite{coag}, pair annihilation
$A+A \ra 0$ \cite{pair,mer,HS} or two-species annihilation $A+B \ra \emptyset$
\cite{2p,ben}. Formulated as a lattice model, $\emptyset$
corresponds to a vacancy on a site of the lattice and  particles
represented by particle occupation numbers may hop in the lattice
($A\emptyset \ra \emptyset A$) and take part in the reactions.
Such lattice systems are, in general, difficult to
treat by rigorous means and correspondingly, considering the vast amount of
such models, relatively few exact results are known.

Over the past few years the formulation of stochastic processes in terms of
quantum spin systems has turned to be a convenient tool in the study of
non-equilibrium lattice problems (see e.g.
\cite{mer,HS,scsa,GS,sch,asym,adhr,Rit} and
references therein). A paradigmatic example is the
representation of symmetric diffusion of hard core particles (known as the
symmetric exclusion process \cite{Spi,Lig}) by the spin-1/2 Heisenberg model
(see \cite{scsa} for a detailed discussion). But also
asymmetric hopping \cite{HS,GS,sch,asym}, multimer processes \cite{mer} and
reaction-diffusion processes \cite{HS,adhr,Rit} have been similarly
represented.
By these means, standard techniques for quantum spin systems such as spin
wave theory \cite{mer,spinwave}, Bethe ansatz and related algebraic techniques
\cite{HS,GS,sch,adhr,Rit,BA}, global symmetries
\cite{scsa,asym} and Goldstone broken
symmetry arguments \cite{mer} have given many new results for stochastic
systems.

An intriguing feature of these models is that even though these are
interacting many-particle systems, some of them are known to give rise to
closed systems of differential-difference equations for time-dependent
correlation functions and exact results have been obtained.
In one possible scenario the time derivative of a $k$-point correlation
function does not involve higher order correlators,
or, in other words, one obtains a closed set of not more than $k$ coupled,
linear, differential-difference equations. Some well-known examples of
single-species processes where this happens are the symmetric exclusion process
and symmetric partial exclusion process describing diffusion of
particles on a lattice in any dimension \cite{scsa,Spi,Lig}, the
asymmetric exclusion process describing driven diffusion in one dimension
\cite{asym} or the voter model
describing annihilation (death) processes and decoagulation \cite{Lig}
(for details see below). In another scenario, certain subsets of correlation
function decouple from each other. This happens e.g. in the Glauber model
\cite{Glauber}, in random sequential
dimer deposition \cite{evans} or in the generalized models studied in
Ref. \cite{prs}.

This observation raises a number of questions, the most obvious one being
whether there is a classification of these processes, i.e., a general criterion
on the reaction and diffusion rates such that the resulting equations for the
correlation functions decouple. Clearly, the answer depends on which
correlation functions one wishes to study. Here we consider density correlation
functions which are the ones one is usually interested in. Other correlation
functions e.g. exponentials of integrated densities \cite{asym} or
particle-string correlation functions \cite{asym,evans,prs} give rise to
other necessary and sufficient equations for the rates leading to decoupled
equations. As we shall show below, the quantum
Hamiltonian formalism that we use throughout this paper is a convenient tool
for the derivation of these equations (\ref{3-8}) for the rates
and it opens the way to a (partial)
physical and mathematical understanding of this phenomenon. This is because
by expressing the process in terms of a quantum Hamiltonian one relates
a $d$-dimensional non-equilibrium problem to a $d+1$-dimensional
non-equilibrium
problem into which one may have (and in the cases discussed here actually
{\em has}) some insight. Here we study only stochastic processes of hard-core
particles with a two-site interaction. The corresponding quantum Hamiltonian
turns out to be that of a  generalized spin-1/2 Heisenberg model. However,
our strategy is easily generalized to many-species models or to models with
interactions involving more than only two sites.

In some of the known cases which are contained in our more general model
the observed decoupling of the correlation functions can be understood in terms
of a dual stochastic process \cite{Spi,Lig}. In the case of the (self dual)
symmetric exclusion process duality relates the time-dependent $k$-point
density correlation function to the process with a $k$-particle initial state.
Because of particle number conservation this is a great simplification
and many exact results have been obtained in this way. Also for other processes
duality may be used to derive new results \cite{Lig} and so one other question
we discuss is the existence of dual processes to those which satisfy the
constraints (\ref{3-8}) on the rates discussed above. The dual processes
that we shall obtain involve additional restriction on the rates arising
from the positivity of the dual rates and conservation of probability in the
dual process. In any case, the dual
process contains only hopping and various annihilation terms, but no
non-zero particle creation rates. This is another way of understanding the
decoupling of the correlation functions from higher order correlators.

One more problem that we shall address, albeit only briefly, is that of the
integrability of the system. From the structure of the equations derived in
Sec.~3 and from the solution of these equations for the one-point function
(i.e., the density profile) obtained in Sec.~6 it becomes apparent that the
system is partially integrable in any space dimension. This means that
some of the equations for the $k$-point functions are
integrable and therefore yield the spectrum of a subspace of the Hamiltonian.
In Sec.~5 we show that the spectrum of $H$ is identical to that
of a spin-1/2 Heisenberg Hamiltonian which, in one dimension, is completely
integrable by the Bethe ansatz. However, the generalized Hamiltonian $H$ is not
related to the Heisenberg Hamiltonian by a similarity transformation. The
integrability of $H$ can not be derived from the usual Baxterization procedure
\cite{Bax} and the algebraic structure underlying the generalized model
remains an open problem.

The paper is organized as follows. In Sec.~2 we give for the benefit of the
reader not familiar with the quantum Hamiltonian formalism a full discussion of
its relation to the usual description of the stochastic process in terms of a
master equation. We also introduce various definitions and relations used
later. We define the problem on a hypercubic lattice in $d$-dimensions with
periodic boundary conditions even though many of the results derived later
are also valid for other lattices or boundary conditions. When appropriate,
this is indicated by an additional explicit remark. Otherwise one should
always think of the periodic hypercubic lattice. In Sec.~3 we derive
the equations for the rates such that one obtains linear, inhomogeneous
differential-difference equations, i.e., equations for the $k$-point
correlation functions  which are decoupled from higher order correlators.
In Sec.~4 we discuss duality and derive the criteria on the existence of a dual
process. This leads also to a number of duality relations for the correlation
functions. In Sec.~5 we discuss the mapping of the stochastic problem to the
Heisenberg quantum Hamiltonian and in Sec.~6 we calculate the time-dependent
density profile
with an arbitrary initial state. In Sec.~7 we summarize the main results and
present some open questions.

\section{Stochastic processes in the quantum Hamiltonian formalism}
\setcounter{equation}{0}

We study one-species exclusion processes in $d$ dimensions, i.e., a system of
particles on a hypercubic lattice with $M$ sites
where each site is either empty or occupied
by at most one particle. The state space of the system is therefore
$X = \{ 0,1 \}^M$ and a given state of the system may be represented by a
configuration $\udl{n} = \{n_1, n_2, \dots , n_M\}$ where $n_i = 0,1$ and
$1 \leq i \leq M$ labels the sites of the lattice.
An alternative possibility is to give the set
$\{ \vec{x}_1, \vec{x}_2,
\dots , \vec{x}_N \}$ of occupied lattice sites. In this notation,
the empty set
represents the empty lattice and $1 \leq N \leq M$ is the total number of
particles in the configuration. $\vec{x}_i=(x^{(1)}_i,
x^{(2)}_i, \dots , x^{(d)}_i)$ is a $d$ component object defining the (integer)
coordinates of the particle in the lattice. When working on finite lattices,
we shall label each space coordinate by an integer
$1 \leq x^{(a)} \leq L^{(a)}$, for
an infinite lattice $x^{(a)} \in {\bf Z}$. For later convenience, it is useful
to introduce also the unit vector in $a$-direction, $\vec{e}^{(a)} = (0^{(1)},
\dots ,
1^{(a)}, \dots , 0^{(d)})$.

The stochastic dynamics of the system may be defined
in terms of a master equation for the probability $f(\udl{n};t)$
of finding the configuration $\{n_1, n_2,
\dots , n_M \}$ at time $t$. We shall use quantum Hamiltonian language
which has proven to be a useful formalism for stochastic processes on lattices.
Each state $\udl{n} \in X$ is represented by a vector
$|\, \udl{n} \,\rangle$ (or $|\, \vec{x}_1, \dots , \vec{x}_N \,\rangle$, with
$|\,0\,\rangle\equiv |\;\;\rangle$ being the empty state) and the probability
distribution is mapped to a state vector
\bel{2-1}
| \, f(t)\, \rangle = \sum_{\udl{n} \in X} f(\udl{n};t)
|\, \udl{n} \,\rangle \htwo .
\ee
The vectors $|\, \udl{n} \,\rangle$
together with the
transposed vectors $\langle \, \udl{n} \, |$ form an orthonormal basis
of $(C^2)^{\otimes M}$ and the time evolution is defined
in terms of a linear 'Hamilton' operator $H$ acting on this space of
dimension $2^M$
\bel{2-2}
\frac{\partial}{\partial t} | \, f(t)\, \rangle =
- H | \, f(t)\, \rangle \htwo .
\ee
A state at time $t=t_0 + \tau$ is therefore given in terms of an initial
state at time $t_0$ by
\bel{2-2a}
| \, f(t_0+\tau \, \rangle = \mbox{e}^{-H\tau }
| \, f(t_0 \, \rangle \htwo .
\ee
{}From (\ref{2-1}) and (\ref{2-2}) and using
$f(\udl{n};t) = \langle \, \udl{n} \, | \, f(t) \, \rangle$
the master equation takes the form
\bel{2-3}
\frac{\partial}{\partial t} f(\udl{n};t) = - \langle \, \udl{n} \, |
H | \, f(t) \, \rangle \htwo .
\ee

Note that
\bel{2-4}
\langle \, s \,|\, f(t) \, \rangle = \sum_{\udl{n} \in X} f(\udl{n};t) = 1
\ee
where
\bel{2-5}
\langle \, s \,| = \sum_{\udl{n} \in X} \langle \, \udl{n} \, |
\ee
which expresses conservation of probability. This implies
$\langle \, s \,| H = 0$ for any stochastic process.
The right eigenvector(s) of $H$ with eigenvalue
$E_0 = 0$ and normalized according to (\ref{2-4}) is (are) the
steady state(s) of the stochastic process.
In general $H$ is not symmetric which means that the rate
$w(\udl{n};\udl{n}') = - \langle \, \udl{n} \, | H |\, \udl{n}' \,\rangle$
with which a configuration $\udl{n}'$ switches to a configuration $\udl{n}$
is not, in general, equal to its reverse rate $w(\udl{n}';\udl{n})$.
As a result the stationary distribution(s) $S(\udl{n})$ may be highly
non-trivial.
The real part of all eigenvalues of $H$ is larger or equal to zero.

Average values $\langle \, Q \, \rangle$ are calculated as matrix elements
of suitably chosen operators $Q$
which may be expressed in terms of the usual Pauli matrices
$\sigma^{x,y,z}_{\vec{j}}$
acting on site $\vec{j}$.
A complete set of observables are the occupation numbers $n_{\vec{j}}=0,1$.
Defining projection {\em operators} on states with a particle on site $\vec{j}$
of
the chain as
\bel{2-7}
n_{\vec{j}} = \frac{1}{2} \left( 1 - \sigma^z_{\vec{j}} \right) =
\left(
\ba{cc}
0 & 0 \vtwo \\
0 & 1
\ea
\right)_{\vec{j}}
\ee
one finds that the average density of particles at
site $\vec{j}$ is given by $\langle \, n_{\vec{j}} \, \rangle =
\langle \, s \,| n_{\vec{j}} | \, f(t) \, \rangle$.
Correlation functions
$\langle \, n_{\vec{x}_1} \cdots n_{\vec{x}_k} \, \rangle$, i.e., the
probabilities of finding particles on the set of sites
$\{\vec{x}_1, \dots ,\vec{x}_k\}$,
are computed analogously. Note that in ordinary quantum mechanics average
values would be taken as matrix elements between normalized
eigenstates of $H$, i.e.,
$\langle \, Q \, \rangle = \langle \, k \, | Q | \, k \, \rangle$ whereas
here an average value is the quantity $\langle \, s \, | Q | \, f \, \rangle$
where $| \, f \, \rangle$ is (in general) not an eigenstate, but a state
with real coefficients $0 \leq f(\udl{n};t) \leq 1$ (\ref{2-1}) in the basis
spanned by the set $\{ | \udl{n} \rangle \}$ and normalized such that
$\langle \, s \, | \, f \, \rangle=1$.

For later convenience we also introduce the operators
$s^{\pm}_{\vec{j}} = (\sigma^x_{\vec{j}} \pm i \sigma^y_{\vec{j}})/2$.
In our convention
\bel{2-8}
s^-_{\vec{j}} =
\left(
\ba{cc}
0 & 0 \vtwo \\
1 & 0
\ea
\right)_{\vec{j}}
\ee
creates a particle at site $\vec{j}$ when acting to the right, while
\bel{2-9}
s^+_{\vec{j}} =
\left(
\ba{cc}
0 & 1 \vtwo \\
0 & 0
\ea
\right)_{\vec{j}}
\ee
annihilates a particle at site $\vec{j}$. Note that
\bel{2-10}
\langle \, s \,| s^+_{\vec{j}} = \langle \, s \,| n_{\vec{j}}
\hfour \mbox{and}  \hfour
\langle \, s \,| s^-_{\vec{j}} = \langle \, s \,| (1-n_{\vec{j}} ) \htwo .
\ee
Introducing the ladder operator $S^{\pm} = \sum_{\vec{j}} s^{\pm}_{\vec{j}}$
one may write
\bel{2-10a}
\langle \, s \,| = \langle \, 0 \,| \, \mbox{e}^{S^+} \htwo .
\ee
Using the commutation relations for the Pauli matrices then yields
(\ref{2-10}).

Now we are in a position to define the stochastic processes we intend to
study by a quantum Hamiltonian $H$. We define
\bel{2-11}
H = \xi \sum_{\vec{j}} \sum_{a=1}^{d} u_{\vec{j}}^{(a)}
\ee
with the nearest neighbour reaction matrices
\bel{2-12}
 u_{\vec{j}}^{(a)} = - \left(
\ba{cccc}
a_{11} & a_{12} & a_{13} & a_{14} \vtwo \\
a_{21} & a_{22} & a_{23} & a_{24} \vtwo \\
a_{31} & a_{32} & a_{33} & a_{34} \vtwo \\
a_{41} & a_{42} & a_{43} & a_{44}
\ea \right)_{(\vec{j};a)}
\ee
acting on nearest neighbour sites $\vec{j}$ and $\vec{j}+\vec{e}^{(a)}$.
The sum runs over the whole lattice
and we take periodic boundary conditions
in all space directions, i.e.,
$u_{(j^{(1)},\dots ,L^{(a)}, \dots ,j^{(d)})}^{(a)}$ acts on sites
$(j^{(1)},\dots ,L^{(a)}, \dots ,j^{(d)})$ and
$(j^{(1)},\dots ,1^{(a)}, \dots ,j^{(d)})$.
The normalization $\xi$ sets the time scale and is for the purposes of this
paper of no particular interest.
The diagonal elements $a_{kk}$ of $u_{\vec{j}}^{(a)}$
satisfy
\bel{2-13}
a_{kk}=-\sum_{\stackrel{k'=1}{k'\neq k}}^{4} a_{k'k}
\ee
which is imposed by conservation of probability. This implies
\bel{2-14}
\langle \, s \,| u_{\vec{j}}^{(a)} = 0 \honecm \forall \, \vec{j},a
\ee
In order to keep the interpretation of $H$ as defining a stochastic process,
(\ref{2-13}) has to be supplemented by the condition $a_{kk'} \geq 0$ for
the off diagonal matrix elements $k\neq k'$.

The processes decribed by $H$ are reactions changing the configurations
on two nearest neighbour sites. A configuration
$\{n_{\vec{j}},
n_{\vec{j}+\vec{e}^{(a)}}\}$ changes into configurations  $\{n_{\vec{j}}',
n_{\vec{j}+\vec{e}^{(a)}}'\}$
with rates $a_{kk'}$
as follows:
\bel{2-15}
\ba{lll}
\{0,0\} & \ra & a_{21} \{0,1\} + a_{31} \{1,0\} + a_{41} \{1,1\}
\htwo \mbox{(birth/pair creation)} \vtwo \\

\{0,1\} & \ra & a_{12} \{0,0\} + a_{32} \{1,0\} + a_{42} \{1,1\}
\htwo \mbox{(death/diffusion/decoagulation)}\vtwo \\

\{1,0\} & \ra & a_{13} \{0,0\} + a_{23} \{0,1\} + a_{43} \{1,1\}
\htwo \mbox{(death/diffusion/decoagulation)}\vtwo \\

\{1,1\} & \ra & a_{14} \{0,0\} + a_{24} \{0,1\} + a_{34} \{1,0\}
\htwo \mbox{(pair annihilation/coagulation)} \htwo .
\ea
\ee
These processes take place with equal rates everywhere in the lattice.

This generalized nearest neighbour exclusion process includes many well-known
processes such as the asymmetric exclusion process \cite{Lig}
(with hopping rates $a_{23},a_{32} \neq 0$, all other rates 0),
the voter model \cite{Lig} (with death rates and decoagulation rates
$a_{12}=a_{13}=a_{42}=a_{43}\neq 0$) or Glauber dynamics \cite{Glauber}
($a_{23}+a_{32}=a_{14}+a_{41}\neq 0$). Altogether there are independent 12
parameters,
one of which is trivial as one may always change the normalization $\xi$
without changing the physical properties of the system. We shall set
$\xi=1$ throughout the paper.

\section{Equal-time correlation functions}
\setcounter{equation}{0}

The equal-time $k$-point correlation function satisfies the equation
\bel{3-1}
\frac{\partial}{\partial t}
\langle \, n_{\vec{x}_1} \cdots n_{\vec{x}_k} \, \rangle
=  - \sum_{a=1}^{d} \sum_{\vec{j} \in C^{(a)}}
\langle \, n_{\vec{x}_1} \cdots n_{\vec{x}_k} u_{\vec{j}}^{(a)} \, \rangle
\ee
where, owing to the property (\ref{2-14}) of the two-site reaction matrix
$u_{\vec{j}}^{(a)}$, the sum over $\vec{j}$ does not run over the whole
lattice, but only over the union
$C^{(a)}$ of the set of sites $\{\vec{x}_1,\dots,\vec{x}_k\}$
with the set of their nearest neighbours
$\{ \vec{x_1}-\vec{e}^{(a)}, \dots , \vec{x_k}-\vec{e}^{(a)} \}$.
It is important to realize that
since each $u_{\vec{j}}^{(a)}$ acts non-trivially
only on two sites, the r.h.s. of (\ref{3-1})
involves only $(k-2)$-point functions, $(k-1)$-point functions,
$k$-point functions and $(k+1)$-point functions. This can be seen as follows:
Suppose one of the $\vec{x}_i \in \{\vec{x}_1,\dots,\vec{x}_k\}$ (say
$\vec{x}_k$) is equal to $\vec{j}+\vec{e}^{(a)}$. Using (\ref{2-10}) one finds
\bel{3-2}
\ba{lll}
\langle \, n_{\vec{x}_1} \cdots n_{\vec{x}_{k-1}} (n_{\vec{x}_k}
           u_{\vec{x}_k-\vec{e}^{(a)}}^{(a)}) \, \rangle
& = & A_1 \langle \, n_{\vec{x}_1} \cdots n_{\vec{x}_{k-1}} \, \rangle \vtwo \\
&   & + \, B_1 \langle \, n_{\vec{x}_1} \cdots n_{\vec{x}_{k-1}}
          n_{\vec{x}_k-\vec{e}^{(a)}} \, \rangle \vtwo \\
&   & - \, C_1 \langle \, n_{\vec{x}_1} \cdots n_{\vec{x}_{k-1}}
          n_{\vec{x}_k} \, \rangle \vtwo \\
&  & + \, D_1 \langle \, n_{\vec{x}_1} \cdots n_{\vec{x}_{k-1}}
          n_{\vec{x}_k} n_{\vec{x}_k-\vec{e}^{(a)}} \, \rangle
\ea
\ee
with
\bel{3-3}
\ba{llllll}
A_1 & = & a_{21} + a_{41}       &
B_1 & = & a_{23} + a_{43} - a_{21} - a_{41} \vtwo \\
C_1 & = & a_{12} + a_{32} + a_{21} + a_{41} \hfour
                 & D_1 & = & C_1 - a_{23} - a_{43} - a_{14} - a_{34} \htwo .
\ea
\ee

A similar result arises if one
of the $\vec{x}_i$ (again, without loss of generality $\vec{x}_k$)
is equal to $\vec{j}$:
\bel{3-4}
\ba{lll}
\langle \, n_{\vec{x}_1} \cdots n_{\vec{x}_{k-1}} (n_{\vec{x}_k}
           u_{\vec{x}_k}^{(a)}) \, \rangle
& = & A_2 \langle \, n_{\vec{x}_1} \cdots n_{\vec{x}_{k-1}} \, \rangle \vtwo \\
&   & + \, B_2 \langle \, n_{\vec{x}_1} \cdots n_{\vec{x}_{k-1}}
          n_{\vec{x}_k+\vec{e}^{(a)}} \, \rangle \vtwo \\
&   & - \, C_2 \langle \, n_{\vec{x}_1} \cdots n_{\vec{x}_{k-1}}
          n_{\vec{x}_k} \, \rangle \vtwo \\
&   & + \, D_2 \langle \, n_{\vec{x}_1} \cdots n_{\vec{x}_{k-1}}
          n_{\vec{x}_k} n_{\vec{x}_k+\vec{e}^{(a)}} \, \rangle
\ea
\ee
with
\bel{3-5}
\ba{llllll}
A_2 & = & a_{31} + a_{41}       &
B_2 & = & a_{32} + a_{42} - a_{31} - a_{41}\vtwo \\
C_2 & = & a_{13} + a_{23} + a_{31} + a_{41} \hfour
                 & D_2 & = & C_2 - a_{32} - a_{42} - a_{14} - a_{24} \htwo .
\ea
\ee
The r.h.s. of
(\ref{3-2}) and (\ref{3-4}) consist only of  $(k-1)$-point functions,
$k$-point functions and $(k+1)$-point functions.

If two of the $\vec{x}_i$ are nearest neighbours in the lattice, e.g.
$\vec{x}_{k-1}=\vec{x}_k-\vec{e}^{(a)}=\vec{j}$, then the action of
$u_{\vec{j}}^{(a)}$  yields
\bel{3-6}
\ba{lll}
\langle\,n_{\vec{x}_1}\cdots n_{\vec{x}_{k-2}}
     (n_{\vec{x}_{k-1}}n_{\vec{x}_{k-1}+\vec{e}^{(a)}}
     u_{\vec{x}_{k-1}}^{(a)}) \, \rangle
& = & A_3 \langle \, n_{\vec{x}_1} \cdots n_{\vec{x}_{k-2}} \, \rangle \vtwo \\
&   & + \, B_3 \langle \, n_{\vec{x}_1} \cdots n_{\vec{x}_{k-2}}
          n_{\vec{x}_{k-1}+\vec{e}^{(a)}} \, \rangle \vtwo \\
&   & + \, D_3 \langle \, n_{\vec{x}_1} \cdots n_{\vec{x}_{k-2}}
          n_{\vec{x}_{k-1}} \, \rangle \vtwo \\
&   & - \, C_3 \langle \, n_{\vec{x}_1} \cdots n_{\vec{x}_{k-2}}
          n_{\vec{x}_{k-1}} n_{\vec{x}_{k-1}+\vec{e}^{(a)}} \, \rangle \htwo .
\ea
\ee
with
\bel{3-7}
\ba{llllll}
A_3 & = & a_{41}          & B_3 & = & a_{42} - a_{41} \vtwo \\
D_3 & = & a_{43} - a_{41} \hfour
      & C_3 & = & a_{14} + a_{24} + a_{34} + a_{42} + a_{43} - a_{41} \htwo .
\ea
\ee
The r.h.s. of (\ref{3-6}) consists only of  $(k-2)$-point functions,
$(k-1)$-point functions and $k$-point functions.

If $D_1 \neq 0$ or $D_2 \neq 0$ the time derivative (\ref{3-1}) of the
$k$-point correlation function gives rise to
a set of $M$ coupled differential-difference equations involving all
$k$-point correlators.\footnote{In $N$-particle systems with
particle number conservation the hierachy breaks off at $k=N\leq M$.}
Solutions to
such a set of equations have been found in some special cases where subsets
of these equations decouple \cite{prs},
but there is no general solution. On the
other hand, if $D_1=D_2=0$, i.e., for
\bel{3-8}
\ba{lll}
a_{34} & = & a_{21} + a_{41} + a_{12} + a_{32}
             - a_{23} - a_{43} - a_{14} \vtwo \\
a_{24} & = & a_{31} + a_{41} + a_{13} + a_{23}
             - a_{32} - a_{42} - a_{14} \htwo ,
\ea
\ee
the problem simplifies considerably as one has a closed
system of only $k$ equations. In this case Eq. (\ref{3-1})
may be regarded as an inhomogeneous,
linear differential-difference equation for the $k$-point function with
$(k-1)$-point and $(k-2)$-point correlators as inhomogeneities.

Note that one may also study correlation functions of the operators
$\tilde{n}=n-\alpha$ with an arbitrary constant $\alpha$. One obtains
again a closed system of $k$ equations for the $k$-point correlation
function if $D_1=D_2=0$, but with new constants
\bel{3-8a}
\tilde{A}_1 \; = \; A_1 + \alpha (B_1-C_1) \; , \hsix
\tilde{A}_1 \; = \; A_1 + \alpha (B_1-C_1)
\ee
and
\bel{3-8b}
\ba{lll}
\tilde{A}_3 & = & A_3 + \alpha (B_3+D_3) - \alpha^2 C_3 \vtwo \\
\tilde{B}_3 & = & B_3 - \alpha (C_3+B_2-C_1) \vtwo \\
\tilde{D}_3 & = & B_3 - \alpha (C_3+B_1-C_2)
\ea
\ee
in Eqs. (\ref{3-2}), (\ref{3-4}), (\ref{3-6}). $B_{1,2}$ and $C_{1,2}$ do not
change. In particular,
the inhomogeneity arising from $A_1,A_2 \neq 0$ in the one-point function
can be removed by  taking
$\alpha=\rho$ with
\bel{3-9}
\rho \, = \, \frac{ 2 a_{41} + a_{21} + a_{31} }
{ 2 a_{41} + a_{21} + a_{31} + 2 a_{14} + a_{24} + a_{34}} \, = \,
\frac{A_1 + A_2}{C_1 + C_2 - B_1 - B_2} \htwo .
\ee
With this choice one has $\tilde{A}_1+\tilde{A}_2=0$.
The differential-difference equation for higher order correlation functions
have then
inhomogeneous terms proportional to $\tilde{A}_3$ (coupling to
$(k-2)$-point functions) and $\tilde{B}_3'=\tilde{B}_3+\tilde{A}_1$,
$\tilde{D}_3'=\tilde{D}_3+\tilde{A}_2$ (coupling to $(k-1)$-point functions.
We conclude: \vtwo \\
{\em  Eq. (\ref{3-1}) becomes
a closed, inhomogeneous, linear differential-difference equation in one
(continous) time coordinate and $d\cdot k$ (discrete) space coordinates
on a 10-parameter submanifold defined by Eqs. (\ref{3-8})
of the 12-parameter model.}\vtwo \\
One may add the remark that this differential-difference equation
becomes homogeneous (i.e., contains no $(k-1)$-point and $(k-2)$-point
correlation functions) on a 7-parameter submanifold defined by
$\tilde{A}_3=\tilde{B}_3'=\tilde{D}_3'=0$. From the derivation
presented above it is obvious that this result is easy
to generalize to other lattices and interactions. With $D_1=D_2=0$
Eq. (\ref{3-1}) is a closed set of $k$ equations
independent of the dimensionality of the system or of the kind of lattice on
which the model is defined. Furthermore, the two sites on which the reaction
matrix $u$ acts nontrivially are not even required to be nearest neighbours.
The result remains true for arbitrary long-range interactions with
reaction matrices $u_{\vec{x},\vec{y}}$ where $\vec{x}$ and $\vec{y}$ are
any two points on the lattice. Finally,
it is also not necessary to keep the reaction rates $a_{kk'}$
space-independent as long as (\ref{3-8}) is satisfied for each reaction
matrix $u_{\vec{x},\vec{y}}$. For the decoupling from lower order correlators
a stronger condition is necessary for this general case. Besides
$D_1=D_2=0$ one needs
$\tilde{A}_1=\tilde{A}_2=\tilde{A}_3=\tilde{B}_3=\tilde{D}_3=0$, i.e., one is
left with a five-parameter space only.

We demonstrate this result for the one-dimensional case. In one dimension,
Eq. (\ref{3-1}) for the one-point function becomes
\bel{3-10}
\frac{\partial}{\partial t} \langle \, n_x \, \rangle = -
\langle \, n_x \, (u_{x-1} + u_x) \, \rangle
\ee
where we have set $\xi=1$ and dropped the coordinate index $a$
in $u_{\vec{j}}^{(a)}$. This is easy to calculate and one finds
\bel{3-11}
\ba{lll}
\disp \frac{\partial}{\partial t} \langle \, n_x \, \rangle & = &
A_1 + A_2 \vtwo \\
 & & + B_1 \exval{n_{x-1}} - (C_1 + C_2) \exval{n_x} + B_2 \exval{n_{x+1}}
\vtwo \\
 & & + D_1 \exval{n_{x-1}\,n_x} + D_2 \exval{n_{x}\,n_{x+1}}
\ea
\ee
where the constants with index 1 and index 2
arise from the action of $u_{x-1}$ and $u_x$ respectively.
One sees that if $D_1=D_2=0$ (\ref{3-11}) becomes an inhomogeneous, linear
differential-difference equation. Introducing $\tilde{n}$ with $\alpha=\rho$
as defined in (\ref{3-9}) leads to the homogeneous equation
\bel{3-12}
\frac{\partial}{\partial t} \langle \, \tilde{n}_x \, \rangle =
B_1 \exval{\tilde{n}_{x-1}} + B_2 \exval{\tilde{n}_{x+1}} -
(C_1 + C_2) \exval{\tilde{n}_x}  \htwo .
\ee
For the two-point function one obtains
\bel{3-13}
\ba{lll}
\disp \frac{\partial}{\partial t} \langle \, n_{x} n_{y} \, \rangle & = & -
\langle\, n_{x} (u_{x-1} + u_{x}) n_y \,\rangle -
\langle\, n_{x} n_y (u_{y-1} + u_{y}) \,\rangle \vtwo \\
 & = & (A_1+A_2)(\exval{n_x}+\exval{n_y}) \vtwo \\
 &   & B_1 (\exval{n_{x-1} n_y}+\exval{n_x n_{y-1}}) +
       B_2 (\exval{n_{x+1} n_y}+\exval{n_x n_{y+1}}) \vtwo \\
 &   & -2(C_1+C_2) \langle \, n_{x} n_{y} \, \rangle
\ea
\ee
if $x$ and $y$ are not nearest neighbours and
\bel{3-14}
\ba{lll}
\disp \frac{\partial}{\partial t} \langle \, n_{x} n_{x+1} \, \rangle & = & -
\langle\, n_{x} n_{x+1} (u_{x-1} + u_{x} + u_{x+1}) \,\rangle
\vtwo \\
 & = & A_3 + (A_2+D_3)\exval{n_x} +
                (A_1+B_3)\exval{n_{x+1}} \vtwo \\
 &   & + B_1 \exval{n_{x-1}n_{x+1}} + B_2 \exval{n_{x}n_{x+2}} \vtwo \\
 &   & - (C_1 + C_2 + C_3)
       \exval{n_{x}n_{x+1}} \htwo .
\ea
\ee
for the nearest neighbour correlator. Similar equations are obtained for
correlators involving $\tilde{n}$.

For $D_1=D_2=0$ and the special choice $a_{k1}=a_{4k}=0$
(no birth, pair creation and decoagulation) one has $A_i=B_3=D_3=0$
and (\ref{3-14}) simplifies to the completely decoupled, homogeneous equation
\bel{3-15}
\ba{lll}
\disp \frac{\partial}{\partial t} \langle \, n_{x} n_{x+1} \, \rangle
 & = & B_1 \exval{n_{x-1}n_{x+1}} + B_2 \exval{n_{x}n_{x+2}} \vtwo \\
 &   & - (C_1 + C_2 + C_3)
       \exval{n_{x}n_{x+1}} \htwo .
\ea
\ee
In this case,
time derivatives of higher $k$-point correlation functions also decouple
completely, i.e., involve only $k$-point correlation functions.

\section{Dual processes}
\setcounter{equation}{0}

The fact that the time-derivative of the $k$-point correlation functions may
give rise to a closed set of equations is reminiscent of the duality relations
e.g. for the symmetric exclusion process \cite{Spi,Lig} where self-duality
is indeed just an expression of this fact. One may therefore ask whether
the closure of the equations for the $k$-point functions is equivalent to
the existence of some dual process.

Before we discuss this question we would like to remind the reader of
the meaning of self-duality (of the symmetric exclusion process) in the
operator language used in this paper.
Let us assume that initially $N$ particles are located on a set of sites
$A_N=\{\vec{y}_1,\dots ,\vec{y}_N\}$ represented by a vector
$| \, A_N \, \rangle = | \, \vec{y}_1,\dots ,\vec{y}_N \, \rangle$.
We want to compute the probability
$\langle \, n_{\vec{x}_1} \cdots n_{\vec{x}_k} \, \rangle_{A_N}$ of finding
(any) $k$ particles on sites $B_k=\{\vec{x}_1, \dots , \vec{x}_k\}$,
at time $t$. The duality relations
state \cite{Spi,Lig,scsa}
\bel{4-1}
\langle \, n_{x_1} \cdots n_{x_k} \, \rangle_{A_N} =
\sum_{B_k' \subset A_N} \langle \, n_{x_1'} \cdots n_{x_k'} \,
\rangle_{B_k} \htwo.
\ee
In this expression the sum runs over all sets
$B_k'=\{\vec{x}_1', \dots , \vec{x}_k'\}$
which are contained in the set $A_N$, i.e.,
the $k$-point correlation function
$\langle \, n_{\vec{x}_1} \cdots n_{\vec{x}_k} \rangle_{A_N}$ of the
$N$-particle system is given by sums of $k$-particle
correlation functions (we assume $k \leq N$). Using (\ref{2-10}),
(\ref{2-10a}) and
the fact that $H$ for the symmetric exclusion process is symmetric and
$SU(2)$ invariant (i.e., commutes with $S^+$),
the duality relations (\ref{4-1}) can be derived as follows
\cite{scsa}:
\bea
\langle \, n_{\vec{x}_1} \cdots n_{\vec{x}_k} \rangle_{A_N} & = &
\langle \, s\, | n_{\vec{x}_1} \cdots n_{\vec{x}_k} \mbox{e}^{-Ht}
| \, A_N \, \rangle \nonumber \\
& = &  \sum_{\udl{n}}
\langle \, 0\, | s^+_{\vec{x}_1} \cdots s^+_{\vec{x}_k} \mbox{e}^{-Ht}
| \, \udl{n} \, \rangle\langle \, \udl{n} \, |
\mbox{e}^{S^+} | \, A_N \, \rangle \nonumber \\
\label{4-2}
& = &  \sum_{\udl{n}} \langle \, A_N \, |
\mbox{e}^{S^-} | \, \udl{n} \, \rangle
\langle \, \vec{x}_1, \dots , \vec{x}_k \, | \mbox{e}^{-Ht}
| \, \udl{n} \, \rangle \\
& = &  \sum_{B_k' \subset A_N} \langle \, \vec{x}_1', \dots , \vec{x}_k' \, |
\mbox{e}^{-H^T t} | \, B_k \, \rangle
\nonumber \\
& = &  \sum_{B_k' \subset A_N}
\langle \, s \, | n_{\vec{x}_1'} \cdots n_{\vec{x}_k'}
\mbox{e}^{-Ht} | \, B_k \, \rangle \nonumber \htwo .
\eea
where $| \, B_k \, \rangle = | \vec{x}_1, \dots , \vec{x}_k \rangle$.
Because of particle number conservation we have substituted
$\langle \, \vec{x}_1', \dots , \vec{x}_k' \, |$ in the last line by
$\langle \, s \, | n_{\vec{x}_1'} \cdots n_{\vec{x}_k'}$. In other
words, the averaging is performed over all $k$-particle states such that
the sets $B_k'$ of occupied sites are contained in the set $A_N$ of initially
occupied sites. These sets $B_k'$ arise from the matrix element
$\langle \, A_N \, |\mbox{e}^{S^-} | \, \udl{n} \, \rangle$ together with
particle number conservation.

Generally we define a duality relation by
\bel{4-3}
\langle \, s\, | Q \mbox{e}^{-Ht} | \, A \, \rangle =
\langle \, s\, | Q' \mbox{e}^{-\tilde{H} t} | \, A' \, \rangle
\ee
where $Q$ and $Q'$ are some functions of the projection operators $n_{\vec{x}}$
and $| \, A \, \rangle$ and $| \, A' \, \rangle$ are initial states.
The dual process $\tilde{H}$ is obtained by taking the transposed matrix $H^T$
of the time evolution operator and performing some suitably chosen similarity
transformation $V$ such that
\bel{4-4}
\tilde{H} = V \, H^T \,V^{-1}
\ee
indeed defines a stochastic process. The observable $Q'$ and the initial
condition $| \, A' \, \rangle$ are then given by
\bel{4-5}
\langle \, s\, | Q' = \langle \, A \, | V^{-1} \htwo , \hsix
| \, A' \, \rangle = V Q | \, s \, \rangle
\ee
where $| \, \dots \, \rangle=(\langle \, \dots \, |)^T$. By taking
$V=\exp{(-S^-)}$ and $Q=n_{\vec{x}_1}\cdots n_{\vec{x}_k}$
one recovers (\ref{4-1}) with $\tilde{H}=H$.

In this kind of duality the points $\{ \vec{x}_1, \dots , \vec{x}_k\}$ occuring
in the correlator are mapped to an initial state with particles occupying sites
$\{\vec{x}_1, \dots , \vec{x}_k\}$. The dual process may be considered as a
process describing the time evolution of particles on these points.
$\tilde{H}=H$ means that the symmetric exclusion process is self-dual.

After this reminder we are in a position to formulate the problem more
specificly: We have seen in the previous section that as in the symmetric
exclusion process the time derivative of the $k$-point function of the
generalized model does not involve higher
correlators. The transformation $V=\exp{(-S^-)}$ relates the
$k$-point correlator to a $k$ particle initial state. We therefore ask
the question whether the dual operator $\tilde{H}$ (\ref{4-4})
with this particular $V$ defines a stochastic process on the 10-parameter
manifold $D_1=D_2=0$.
In order to answer this question one has to check whether the transformed
dual rates $\tilde{a}_{kl}$ satisfy (\ref{2-13}) (guaranteeing conservation of
probability)
and the condition of positivity $\tilde{a}_{kl}\geq 0$ for $k\neq l$.

A short calculation gives for the dual matrices
\bel{4-6}
 \tilde{u}_{\vec{j}}^{(a)} = - \left(
\ba{rrrr}
0 &   A_1 &   A_2 &   A_3 \vtwo \\
0 & - C_1 &   B_2 &   B_3 \vtwo \\
0 &   B_1 & - C_2 &   D_3 \vtwo \\
0 &   0   &   0   & - C_3
\ea \right)_{(\vec{j};a)}
\ee
with the quantities $A_i,B_i,C_i$ defined by (\ref{3-3}), (\ref{3-5}) and
(\ref{3-7}) respectively. Using positivity and conservation of probability
the conditions for the existence of a dual process can be
read off, namely $A_i,B_i,D_3\geq 0$ and $A_1+B_1-C_1=A_2+B_2-C_2=
A_3+B_3+D_3-C_3=0$. Note that the condition
$C_3=a_{42}+a_{43}-a_{41}+a_{14}+a_{24}+a_{34}=A_3
+B_3+D_3=a_{42}+a_{43}-a_{41}$
implies $a_{k4}=0 \, \forall \, k$ because of the positivity
of the original rates.

The dual process $\tilde{H}$ has only hopping terms $B_1$ and $B_2$
and various non-vanishing annihilation rates, but no particles are
created. Thus the duality relations (\ref{4-3}) with
$Q=n_{\vec{x}_1} \cdots n_{\vec{x}_k}$ read
\bea
\langle \, n_{\vec{x}_1} \cdots n_{\vec{x}_k} \rangle_{A_N}
 & = & \disp N \langle \, 0 \, | \mbox{e}^{-\tilde{H}t} | \, B_k \, \rangle +
\sum_{\vec{y} \in A_N}
\langle \, \vec{y} \, | \mbox{e}^{-\tilde{H}t} | \, B_k \, \rangle \nonumber \\
 &   & \label{4-7} \, + \, \dots  \, +
\sum_{\vec{y}_1, \dots , \vec{y}_k \in A_N}
\langle \,\vec{y}_1, \dots , \vec{y}_k  \, |
\mbox{e}^{-\tilde{H}t} | \, B_k \, \rangle  \\
& = &
\sum_{p=0}^{k} \sum_{B_p' \subset A_N}
\langle \, s \, | \, Q'(p) \, | \, B_k \, \rangle \nonumber
\eea
where $Q'(p) = P_p n_{\vec{y_1}} \cdots n_{\vec{y_p}}$ and $P_p$ is the
projector on $p$-particle states arising from the matrix element
$\langle \, A_N \, |\mbox{e}^{S^-} | \, \udl{n} \, \rangle$. The averaging
extends therefore over
all $p$-particle states with $0 \leq p \leq k$  such that the sets $B_p'$ of
occupied sites are contained in the set $A_N$ of initially occupied
sites. Note that (\ref{4-7}) holds for any choice of the parameters
$A_i,B_i,D_3$, irrespective of whether $\tilde{H}$ defines a stochastic
process or not.

One may define other dual processes involving other operators $Q$.
The main requirement for the kind of dual processes
we are interested in is that the set of sites $B_k$ defined by the
product of projectors $n_{\vec{x}_i}$ translates into an
initial configuration of occupied sites. This feature determined the
transformation $V$. The results of the preceding section indicate that
it might be interesting to study correlation functions of the shifted
density projectors $\tilde{n}_{\vec{x}_i} = n_{\vec{x}_i} - \alpha$.
We introduce the local operator
\bel{4-8}
w_{\vec{j}} = \left(
\ba{cc}
1+\alpha  & \alpha \vtwo \\
 -\alpha  & 1- \alpha
\ea
\right)_{\vec{j}}
\ee
and
\bel{4-9}
W = \prod_{\vec{j}} w_{\vec{j}} \htwo .
\ee
It is easy to check that $\langle \, s \, |w_{\vec{j}}=\langle \, s \, |$ and
$\langle \, s \, |n_{\vec{j}}=\langle \, s \,
|(n_{\vec{j}}-\alpha)w_{\vec{j}}$.
Therefore
\bel{4-10}
\langle \, s \, | \tilde{n}_{\vec{x}_1} \cdots \tilde{n}_{\vec{x}_k} =
\langle \, s \, | n_{\vec{x}_1} \cdots n_{\vec{x}_k} W
\ee
and
\bel{4-11}
\exval{ \tilde{n}_{\vec{x}_k} \cdots \tilde{n}_{\vec{x}_k} }_{A} =
\langle \, s \, | n_{\vec{x}_1} \cdots n_{\vec{x}_k} \,
\mbox{e}^{-H't} \ket{A'}
\ee
with $H'=WHW^{-1}$ and the transformed initial state
$\ket{A'} = W \ket{A}$.
Now one may apply (\ref{4-7}) to (\ref{4-11}). Averaging is now performed
over the weighted set of states $\udl{n}$ with total particle number $p \leq k$
and with weights $f_A(\vec{y}_1, \dots , \vec{y}_k)=
\langle \, A' \, |\mbox{e}^{S^-} | \, \vec{y}_1, \dots , \vec{y}_k \, \rangle$:
\bel{4-12a}
\exval{ \tilde{n}_{\vec{x}_1} \cdots \tilde{n}_{\vec{x}_k} }_{A} =
\sum_{\stackrel{\vec{y}_1, \dots , \vec{y}_k}{0\leq p \leq k}}
f_A(\vec{y}_1, \dots , \vec{y}_k)
\langle \, \vec{y}_1, \dots , \vec{y}_k \, |
\mbox{e}^{-\tilde{H}t} | \, B_k \, \rangle \htwo .
\ee
One obtains the dual time
evolution operator
$\tilde{H}=V(WHW^{-1})^TV^{-1}$ with the doubly transformed dual reaction
matrices
\bel{4-13}
 {\tilde{u}_{\vec{j}}^{(a)}} = - \left(
\ba{rrrr}
0 &   \tilde{A}_1 &   \tilde{A}_2 &   \tilde{A}_3 \vtwo \\
0 & - C_1         &   B_2         &   \tilde{B}_3 \vtwo \\
0 &   B_1         & - C_2         &   \tilde{D}_3 \vtwo \\
0 &   0           &   0           & - C_3
\ea \right)_{(\vec{j};a)}
\ee
where the quantities defined in Eqs. (\ref{3-8a}), (\ref{3-8b}) are used.
Positivity and conservation of probability yield again the conditions on
the existence of the dual process defined by $\tilde{H}$.

We conclude that \vtwo \\
{\em the closure of the differential-difference equations for the
$k$-point density (or shifted density) correlation functions does not, in
general, imply the existence of a dual stochastic process as discussed here.
Additional constraints on the original reaction rates arise in order to
conserve probability and positivity for the dual process.} \vtwo \\
Other duality relations and dual processes may be obtained by considering
other correlation functions \cite{asym} or transformations to other initial
states.

\section{The Heisenberg Hamiltonian}
\setcounter{equation}{0}

In the previous section we have shown that on a 10-dimensional submanifold
of the 12-parameter problem all equal-time correlation functions can be
calculated by solving (in)homogeneous, linear differential-difference
equations. From the solutions to these equations one may obtain the
spectrum of $H$ (\ref{2-11}) by looking for the poles of the Laplace transform
of the correlation function. From the one-point function (\ref{3-1}) or
(\ref{6-1}) below one finds a series of eigenvalues
\bel{5-1}
E(\vec{k}) = \sum_{a=1}^d \left(
B_1 \mbox{e}^{i k_a} + B_2 \mbox{e}^{-i k_a} -C_1 - C_2 \right)
\honecm 0 \leq k_a < 2\pi
\ee
which may be interpreted as non-relativistic, free single-particle excitations.
The full spectrum would be obtained from the solution to all correlation
functions. From (\ref{5-1}) we find that the system is partially integrable
in any number of space dimensions. Partial integrability is known to occur
also in a different, 7 parameter, subspace of the
model in one dimension \cite{prs}, but
there the energies have a more complicated structure.
Note that in the discussion in the previous section the positivity of
the constants $a_{kl}$ was only necessary for the interpretation of $H$ as
generating as stochastic process. The partial integrability of $H$ is ensured
by
the constraints (\ref{3-8}) alone.

In order to get some insight into the physical origin of the one-particle
excitations and of the structure of the equations for the higher order
correlators we study the relationship of the stochastic Hamiltonian
(\ref{2-11}) to the Heisenberg quantum Hamiltonian defined below. This is
motivated by the symmetric exclusion process in which case (\ref{2-11}) is
the Hamiltonian of the isotropic Heisenberg ferromagnet.

In this section we show that \vtwo \\
{\em the spectrum of
$H$ (\ref{2-11}) with the constraint (\ref{3-8}) is identical to the
spectrum of
the Hamiltonian of an anisotropic spin-1/2
Heisenberg quantum Hamiltonian $H_{XXZ}$
in a magnetic field. The behaviour of the $k$-point correlation functions is
determined by the excitations of the $p$-magnon sector where
$0\leq p\leq k$.} \vtwo \\

The ferromagnetic Heisenberg quantum Hamiltonian
$H_{XXZ}$ is of the same form as (\ref{2-11}), but with
matrices
\bel{5-2}
\ba{lll}
 h_{\vec{j}}^{(a)} & = & \disp -\frac{1}{2\gamma}
\left( \sigma^x_{\vec{j}} \sigma^x_{\vec{j}+\vec{e}^{(a)}} +
       \sigma^y_{\vec{j}} \sigma^y_{\vec{j}+\vec{e}^{(a)}} +
       \Delta \sigma^z_{\vec{j}} \sigma^z_{\vec{j}+\vec{e}^{(a)}} +
       \beta_1 \sigma^z_{\vec{j}} + \beta_2 \sigma^z_{\vec{j}+\vec{e}^{(a)}} +
c
\right) \vfour \\
& = & \disp - \left(
\ba{cccc}
   0   &    0   &    0   &    0   \vtwo \\
   0   & h_{22} &    1   &    0   \vtwo \\
   0   &    1   & h_{33} &    0   \vtwo \\
   0   &    0   &    0   & h_{44}
\ea \right)_{(\vec{j};a)} \htwo .
\ea
\ee
where have chosen the normalization
$\gamma=1$ and the constant $c=-(b_1+b_2+\Delta)$ such that
$h_{23}=h_{32}=1$,
$h_{11}=0$, $h_{22}=-(\beta_2+\Delta)$,
$h_{33}=-(\beta_1+\Delta)$, $h_{44}=-\beta_1-\beta_2$, and
$h_{23}=h_{32}=1$.
This Hamiltonian has a continous $U(1)$ symmetry generated by
$S^z=\sum_{\vec{j}} \sigma^z_{\vec{j}}/2$. Hence $H_{XXZ}$ splits into sectors
with fixed $z$-component of the total spin. By identifying spin up at site
$\vec{j}$ with a vacancy and spin down with a particle, this $U(1)$ symmetry
amounts to particle number conservation. The sector with $S^z=M/2-N$
corresponds to the $N$-particle sector.

In order to make contact with the reaction matrices
$u_{\vec{j}}^{(a)}$ we study the dual matrices $\tilde{u}_{\vec{j}}^{(a)}$
(\ref{4-6}) which where obtained from the original reaction matrices by the
similarity transformation $V$ and transposition (\ref{4-4}).
First we renormalize $\tilde{H}$ by a factor $\xi=\sqrt{B_1B_2}$ and
perform another
similarity transformation $\hat{H}=(\Phi \tilde{H} \Phi^{-1})/\xi$
where $\Phi=\exp{[ \sum_{\vec{j}} (\vec{j}\cdot
\vec{\eta})\sigma_{\vec{j}}]}$ and $\vec{\eta}=\eta\sum_{a=1}^d\vec{e}^{(a)}$
with $\eta=\sqrt{B_1/B_2}$.
For $\tilde{A}_1=\tilde{A}_2=\tilde{A}_3=\tilde{B}_3=\tilde{D}_3=0$ this
manipulation yields matrices
\be
\hat{u}_{\vec{j}}^{(a)} =  - \left(
\ba{cccc}
   0   &    0   &    0   &    0   \vtwo \\
   0   & u_{22} &    1   &    0   \vtwo \\
   0   &    1   & u_{33} &    0   \vtwo \\
   0   &    0   &    0   & u_{44}
\ea \right)_{(\vec{j};a)}
\ee
with $\hat{u}_{11}=0$,
$\hat{u}_{22}=C_1/\xi$, $\hat{u}_{33} = C_2/\xi$, $\hat{u}_{44} = C_3/\xi$
$\hat{u}_{32}=\hat{u}_{23}=1$.
These matrices are identical to the Heisenberg matrices (\ref{5-2}) for an
appropriate choice of $\beta_1,\beta_2$ and $\Delta$. Therefore on the
five-parameter
submanifold of the general model on which one obtains
completely decoupled equations for the $k$-point correlation
functions one finds $\hat{H}=H_{XXZ}$.\footnote{The
transformation $\Phi$
relates the symmetric model $h_{23}=h_{32}$ to the asymmetric model
$h_{23}\neq h_{32}$, see \cite{HS,asym,Nol,Sho} for the one-dimensional case.
It induces non-periodic, twisted boundary
conditions, i.e., the constants $\hat{u}_{23}$ and $\hat{u}_{32}$
in the boundary matrices are different from those in the bulk matrices
$\hat{u}_{\vec{j}}^{(a)}$. In so far $H_{XXZ}$ and $\hat{H}$ agree
in general only up to boundary terms.}
The duality transformation turns the state
$\langle \, s \, | \tilde{n}_{\vec{x}_1} \cdots \tilde{n}_{\vec{x}_k}$
(\ref{4-10}) appearing in
the $k$-point correlation function (\ref{4-11}) into a $k$-particle initial
state. Since $H_{XXZ}$ (and the transformed
stochastic Hamiltonian $\hat{H}$)
conserve particle number, one finds that the correlator is  given
by the $k$-magnon excitations of the Heisenberg quantum Hamiltonian.

In order to understand the more general 10 parameter manifold $D_1=D_2=0$
we split the transformed dual reaction matrices into two parts
$\hat{u}_{\vec{j}}^{(a)}= h_{\vec{j}}^{(a)} - {h'}_{\vec{j}}^{(a)}$
with
\bel{5-5}
{h'}_{\vec{j}}^{(a)} = - \left(
\ba{cccc}
0 &   \hat{A}_1 &   \hat{A}_2 &   \hat{A}_3 \vtwo \\
0 &       0     &       0     &   \hat{B}_3 \vtwo \\
0 &       0     &       0     &   \hat{D}_3 \vtwo \\
0 &       0     &       0     &       0
\ea \right)_{(\vec{j};a)}
 \htwo .
\ee
Thus one may write $\hat{H} = H_{XXZ} + H'$. The crucial point is that
$H_{XXZ}$ conserves particle number and may therefore be block
diagonalized in
blocks with fixed $N$. On the other hand, $H'$ connects a block with particle
number $N$ with blocks with particle numbers $N-1$ and $N-2$, but not
with any $N'\geq N$. The whole
matrix $\hat{H}$ has therefore a triangular structure with block matrices
labelled by $N$ arising from $H_{XXZ}$ on the diagonal and matrix elements
resulting from $H'$ on the upper off-diagonal.
The characteristic polynomial of $\hat{H}$ does not depend on these
off-diagonal entries and therefore the characteristic polynomials of the
stochastic Hamiltonians $H$, $\tilde{H}$ (\ref{2-12}),(\ref{4-8}) and the
Heisenberg Hamiltonian $H_{XXZ}$ (\ref{5-2}) are identical.\footnote{
The same argument was used in Ref. \cite{adhr}, but on another submanifold
of the parameter space.}
The full dynamics of the $k$-point correlation function are determined not
only by the eigenvalues of the time evolution operator, but also by the
eigenstates. Since $H'$ annihilates only, eigenvalues and eigenstates with
$p\leq k$ particles contribute to the dynamics. This explains the partial
integrability of the model: To the one-point function $k=1$ only the
one-magnon sector contributes. This is indeed a non-relativistic free
particle. It is interesting to note that
in one dimension, $H_{XXZ}$ is completely integrable. In this case
the eigenvalues may be found from the Bethe ansatz.

\section{The average density}
\setcounter{equation}{0}

In this section we give an application of the results of Sec.~3.
The simple form of Eq. (\ref{3-1}) for the shifted average particle density
on a hypercubic lattice,
\bel{6-1}
\frac{\partial}{\partial t} \exval{\tilde{n}_{\vec{x}}} =
\sum_{a=1}^{d} \left(
B_1 \exval{\tilde{n}_{\vec{x}-\vec{e}^{(a)}}}
+ B_2 \exval{\tilde{n}_{\vec{x}+\vec{e}^{(a)}}}
- (C_1+C_2) \exval{\tilde{n}_{\vec{x}}} \right) \htwo ,
\ee
allows for an explicit integration and thus the extraction of the
critical and non-critical behaviour of the system. We define
\bel{6-2}
C = C_1 + C_2, \hfour D = 2 \sqrt{B_1 B_2},
\hfour {\cal E} = \frac{1}{2d}
\ln{\left( \frac{B_2}{B_1} \right)}
\ee
and the vector $\vec{\cal E} = {\cal E} \sum_{a=1}^d \vec{e}^{(a)}$
and study first the infinite system. In $d$ dimensions the solution
$\tilde{\rho}(\vec{x},t)=\exval{\tilde{n}_{\vec{x}}}$ to
(\ref{6-1}) is in terms of modified Bessel functions $I_n(x)$ given by
\bel{6-3}
\tilde{\rho}(\vec{x},t) = \sum_{\vec{y}} a_{\vec{y}} \;
\mbox{e}^{-dCt + \vec{\cal E} \cdot (\vec{x}-\vec{y})} \prod_{a=1}^d
I_{x_a-y_a}(Dt)
\ee
with initial condition $\tilde{\rho}(\vec{x},0)=a_{\vec{x}}$. With
$a_{\vec{x}}=a\delta_{\vec{x},\vec{y}}$ and
for large times $t$ and large separations $\vec{r}^2=(\vec{x}-\vec{y})^2$ with
$\vec{r}^2/t$ fixed this may be written
\bel{6-4}
\tilde{\rho}(\vec{r},t) = a \mbox{e}^{-d \mu t}
(2\pi D t)^{-d/2} \mbox{e}^{-\frac{(\vec{r}-\vec{\cal E}Dt)^2}{2Dt}}
\ee
and the physical interpretation of the constants becomes apparent.
$D$ plays the role of a diffusion constant, while $\vec{\cal E}$ is a driving
field leading to an average drift velocity $\vec{v}=D\vec{\cal E}$.
{}From this we find that the system satisfies the Einstein relation
\bel{6-5}
\left( \frac{\partial
v^{(a)}}{\partial {\cal E}^{(a)}}\right)_{{\cal E}^{(a)}=0}
 = D \htwo .
\ee
The constant
\bel{6-6}
\mu = C-(1+\frac{{\cal E}^2}{2})D
\ee
is a decay constant. For $\mu=0$ the system is critical with dynamical
exponent $z=2$.

For a finite-size scaling study of the density on a hypercubic lattice
we consider a one-dimensional system with $L$
sites. The solution to (\ref{6-1}) with periodic boundary condition and
initial condition $\tilde{\rho}(x,0)=a\delta_{x,y}$
(with $1 \leq x,y \leq L$ and $r=x-y$) is
\bel{6-7}
\tilde{\rho}(r,t) = \frac{1}{L} \sum_{k=0}^{L-1}
\exp{ \left\{ \frac{2 \pi i k r}{L}[2C-(B_1+B_2)\cos{\frac{2\pi k}{L}}-
i(B_1-B_2)\sin{\frac{2\pi k}{L}}]t \right\} }
\ee
Introducing the scaling variables $u=\pi(r+(B_1-B_2)t)/L$ and
$\tau = 2\pi (B_1+B_2)t/L^2$ and taking the limit $L \ra \infty$
this becomes
\bel{6-8}
\tilde{\rho}(u,\tau) = \frac{1}{L} \mbox{e}^{-\mu't} \, \theta_3(u|i\tau)
\ee
with the decay constant $\mu'=C-B_1-B_2$ and the Jacobi theta-function
\bel{6-9}
 \theta_3(u|i\tau) = \sum_{n=-\infty}^{\infty}
\mbox{e}^{-\pi n^2 \tau + 2 i n u} \htwo .
\ee
Eqs. (\ref{6-7}) and (\ref{6-8}) describe a density distribution
of width $\Delta=B_1+B_2=D\cosh{\cal E}$
with its center at $x=y-vt$ where $v=B_1-B_2=D\sinh{\cal E}$.
The Einstein relation (\ref{6-5}) therefore hold also in the finite system.
Note that when taking the scaling limit we have
implicitly assumed that ${\cal E} = \epsilon/L$ vanishes proportional to
$L^{-1}$ because otherwise $u$ would diverge. This implies $\Delta\approx D$,
$v \approx D{\cal E}$ and $\mu' \approx \mu$. For $\mu'=0$ and
large values of $\tau$, i.e. for times larger than $L^2$, the density
in the moving frame of reference decays exponentially to its constant
value $\tilde{\rho}_{\infty}=a/L$ with decay constant
$2\pi^2D$. For small values of $\tau$ the density decays with a
power law behaviour,
$\tilde{\rho}(0,\tau) = a L^{-1} (2\pi D \tau)^{-1/2}$. This
may be shown using the Poisson resummation formula, but is already
clear from (\ref{6-4}).

Eq. (\ref{6-1}) may be solved for other boundary conditions such as
open boundary conditions with injection and absorption of particles
\cite{rg94}. We do not want to discuss this here.

If $A_1,A_2 \neq 0$ the constants $B_1$ and $B_2$ can be negative. On bipartite
lattices this leads to an alternating positive and negative (relative) density
$\tilde{\rho}(\vec{x},t)$. For other lattices the situation is more complex.

\section{Conclusions}
\setcounter{equation}{0}

We have studied a general reaction diffusion process of hard-core particles
with two-site interactions on a lattice in $d$ dimensions. There are twelve
parameters (\ref{2-15}) for the various reaction and diffusion rates. On a
10 parameter submanifold defined by (\ref{3-8}) of the parameter space the
differential-difference equations (\ref{3-1}) satisfied by the $k$-point
density correlation functions are
inhomogeneous linear equations involving no higher order correlators.
We have mainly considered a $d$-dimensional hypercubic lattice with nearest
neighbour interaction and space-independent rates, but this result holds
for arbitrary lattices with arbitrary two-site interactions as long as
(\ref{3-8}) is satisfied for the interaction between each pair of sites.
On a $d$-dimensional hypercubic lattice with periodic boundary conditions
Eqs. (\ref{3-1}) decouple completely on a 7 parameter submanifold.

It is perhaps worthwhile pointing out that throughout Sections 3 - 5 we have
assumed that
initially $N$ particles are located on a set of sites $\{ \vec{x}_1, \dots ,
\vec{x}_N \}$. All calculations can be easily repeated for an arbitrary,
time-dependent initial distribution. In this way one can get similar results
for two-time correlation functions.

We have shown that in general the decoupling from higher order correlators
does not imply the existence of a dual process for the time evolution of
a $k$-particle initial state as one has in the special case of the symmetric
exclusion process. This remains true only under further assumptions on the
reaction rates arising from conservation of probability and positivity of rates
in the dual process. The dual process to the general 10 parameter model
(if it exists) is a process involving only diffusion and annihilation
of particles (death, decoagulation and pair annihilation), but no creation.

The Hamiltonian (\ref{2-12}) defining the stochastic process is partially
integrable and has the same
spectrum as a spin-1/2 Heisenberg quantum Hamiltonian $H_{XXZ}$
in a magnetic field and
with twisted boundary conditions. This explains the occurence
of the non-relativistic one-particle excitations appearing in the
time evolution of the density profile in terms of one-magnon excitations
of the Heisenberg model. The dynamics of the $k$-point correlators are given by
$p$ magnon states with $p \leq k$.
In one dimension, $H_{XXZ}$ is completely integrable and its spectrum can be
found from the Bethe ansatz. However, it is interesting to note that even
though
for $D_1=D_2=0$ the spectra of $H$ and $H_{XXZ}$ are identical, the matrices
$u_j$ do {\em not}, in general,
satisfy the Hecke algebra relations
$u_j u_{j\pm 1} u_j -u_i= u_{j\pm 1} u_j u_{j\pm 1} - u_{j\pm 1}$,
$[u_i,u_j]=0$ for
$|i-j|\geq 2$ and $u_j^2=\lambda u_j$ \cite{adhr,Rit}. Through Baxterization
\cite{Bax} this would imply the integrability of the model as in the case
of the normal Heisenberg Hamiltonian (\ref{5-2}). This observation hints at
a more general algebraic structure beyond the usual conditions for
integrability in one dimension.

In Sec. 6 we have used our results for the exact calculation of the time
evolution of the density from an arbitrary initial density and analyzed
its finite-size scaling behaviour in the scaling regime $t\approx L^2$
($L$ is the size of the system). An initially sharp peak in the distribution
widens diffusively and moves with a constant average
velocity. It turns out that the model satisfies the
Einstein relation (\ref{6-5})
relating the drift velocity and the diffusion constant. Depending on the
various reaction rates there is a non-critical region with an additional
exponential decay of the amplitude. Thus the time evolution of the
density profile on a $d$-dimensional hypercubic lattice with periodic
boundary conditions depends only on four combinations of the 10 parameters.
For other boundary conditions or lattices more parameters enter and it would
be interesting to study the corresponding lattice effects.
The lattice diffusion constant may be 0 or
even negative. The latter case corresponds to the development of an
alternating structure of the average density on bipartite lattices before
reaching the
constant stationary density.  Another open question is the behaviour
of the density-density correlation function in the general 10 parameter
model. This quantity may be non-trivial even in the steady state as only under
strong restrictions on the rates the steady state is a product measure.

\section*{Acknowledgments}

The author would like to thank V. Rittenberg and H. Spohn for valuable
discussions and the Isaac Newton Institute where part of this work was done
for kind hospitality. Financial support by the SERC is acknowledged.

\bibliographystyle{unsrt}

\end{document}